\documentclass[graybox]{svmult}

\usepackage{mathptmx}       % selects Times Roman as basic font
\usepackage{helvet}         % selects Helvetica as sans-serif font
\usepackage{courier}        % selects Courier as typewriter font
\usepackage{type1cm}        % activate if the above 3 fonts are
\usepackage{makeidx}         % allows index generation
\usepackage{graphicx}        % standard LaTeX graphics tool
\usepackage{multicol}        % used for the two-column index
\usepackage[bottom]{footmisc}% places footnotes at page bottom
\usepackage{cite}
\makeindex             % used for the subject index
\def\be{\begin{equation}}
\def\ee{\end{equation}}
\def\lb{\label}
\def\bea{\begin{eqnarray}}
\def\eea{\end{eqnarray}}
\def\om{\omega}
\def\l{\lambda}

\begin{document}

\title*{Black holes sourced by a massless scalar}
\author{Mariano Cadoni \and  Edgardo Franzin}
\institute{Mariano Cadoni \at Dipartimento di Fisica, Universit\`a di
Cagliari and INFN, Sezione di Cagliari, Cittadella Universitaria, 09042 Monserrato, Italy;
\email{mariano.cadoni@ca.infn.it}
\and Edgardo Franzin \at Dipartimento di Fisica, Universit\`a di Cagliari
and INFN, Sezione di Cagliari, Cittadella Universitaria, 09042 Monserrato, Italy;
\at CENTRA, Departamento de F\'\i{}sica, Instituto Superior T\'ecnico, Universidade de Lisboa,
Avenida Rovisco Pais 1, 1049 Lisboa, Portugal; \email{edgardo.franzin@ca.infn.it}}
% Use the package "url.sty" to avoid problems with
% special characters used in your e-mail or web address

\maketitle

\abstract*{We construct asymptotically flat black hole solutions of Einstein-scalar gravity sourced by a nontrivial scalar field with $1/r$ asymptotic behaviour.  Near the singularity the black hole  behaves as the Janis-Newmann-Winicour-Wyman solution. The hairy black hole  solutions  allow for a consistent  thermodynamical description. At large mass they  have the same thermodynamical behaviour of the Schwarzschild black hole, whereas for small masses they differ substantially from the latter.}

\abstract{We construct asymptotically flat black hole solutions of Einstein-scalar gravity sourced by a nontrivial scalar field with $1/r$ asymptotic behaviour.  Near the singularity the black hole  behaves as the Janis-Newmann-Winicour-Wyman solution. The hairy black hole  solutions  allow for consistent  thermodynamical description. At large mass they  have the same thermodynamical behaviour of the Schwarzschild black hole, whereas for small masses they differ substantially from the latter. }
\section*{Introduction and motivations\label{sec:1}}

Static, spherically symmetric solutions of Einstein gravity  sourced by scalar fields
have played an important role for the development of black hole physics. The simplest
solution of this kind, describing an asymptotically flat (AF) spherically symmetric
solution with no horizon, sourced by a  scalar with vanishing potential  are known since
a long time~\cite{Janis:1968zz,Wyman:1981bd}
and they are called the  Janis-Newmann-Winicour or  Wyman (JNWW) solutions.
Initially, the search for  AF black holes (BHs) with  scalar hair was motivated  by 
the issue of the uniqueness of the Schwarzschild solution and related to the
``old'' no-hair theorems~\cite{Israel:1967wq,Bekenstein:1971hc}, which 
forbid the existence of BHs if the scalar potential is convex or semipositive definite. 

In the early nineties, it was discovered that  low-energy string models
may allow for BH solutions with scalar
hair~\cite{Gibbons:1987ps,Garfinkle:1990qj,Cadoni:1993yt,Monni:1995vu},
but a non-minimal coupling between the scalar and the 
electromagnetic field is required.

In recent times, the quest for hairy BH and black brane (BB) solutions  has been motivated by 
 the application of the AdS/CFT correspondence  to condensed 
matter systems~\cite{Hartnoll:2008vx,Horowitz:2008bn,Hartnoll:2009sz,
Cadoni:2009xm,Cadoni:2011kv,Cadoni:2011nq, Cadoni:2011yj,Cadoni:2012uf,Cadoni:2012ea}.
In holographic applications the scalar  field has a nice interpretation  as
an order parameter triggering symmetry breaking/phase transitions in the dual field  theory.
Indeed, several---both numerical and analytical---asymptotically AdS BH and BB solutions with a scalar hair
have been found in this context~\cite{Hartnoll:2008vx,Horowitz:2008bn,Hartnoll:2009sz,Cadoni:2009xm,Cadoni:2011nq,Cadoni:2011kv,Charmousis:2009xr,Cadoni:2013hna}.

Shifting from AF to asymptotically AdS BHs allows to circumvent standard no-hair theorems
because in AdS the scalar field may have tachyonic excitations without destabilizing the vacuum~\cite{Breitenlohner:1982bm}.
This led to the formulation of ``new'' no-hair theorems~\cite{Hertog:2006rr}  which relate the existence of BHs with scalar hair
with the violation of the positivity energy theorem (PET)~\cite{Witten:1981mf}.

In this note, based on Ref.~\cite{Cadoni:2015gfa}, we will show as  the expertise achieved   in the holographic context  
can be successfully used to find  AF  BH solutions with scalar  hair.
Extension to AF BH is an important issue  because scalar fields play  a crucial role both in gravitational and particle physics.
On the one hand, the experimental discovery of the Higgs particle  at LHC has confirmed the existence of a fundamental scalar particle~\cite{ATLAS:2012ae}.
On the other hand, the observations of the PLANCK satellite give striking confirmation of cosmological inflation driven by scalar field coupled
to gravity~\cite{Planck:2013jfk}. Moreover,  scalar fields  give a  way to describe dark energy.

The main result presented here is that the solution-generating technique developed in the holographic context  in 
Ref.~\cite{Cadoni:2011nq} can be also successfully used to construct AF BH solutions sourced by a scalar field behaving
asymptotically as an harmonic function.

The structure of the paper is as follows.
In Sect.~\ref{sec:2} we review  the solution-generating technique of Ref.~\cite{Cadoni:2011nq}.
In Sect.~\ref{sec:3} we rederive the JNWW solutions and discuss their main features.
The boundary conditions on the scalar field and the corresponding asymptotic behavior of the scalar potential are discussed in Sect.~\ref{sec:4}.
In Sect.~\ref{sec:5} we present our hairy BH solutions.
The thermodynamical behaviour of our solutions is discussed in Sect.~\ref{sec:6}.
Finally, in Sect.~\ref{sec:7} we present our conclusions.

\section{The solution-generating technique\label{sec:2}}

We consider Einstein gravity in four spacetime dimensions  minimally coupled to a scalar $\phi$ ($\mathcal{R}$ is the scalar   curvature), 
\be%
A=\int{}d^4x\sqrt{-g}\left(\mathcal{R} -2 (\partial\phi)^{2}-V
(\phi)\right)\lb{action}
\ee%
and we look for static, spherically  symmetric solutions of the field equations,
\be\label{pmetric}
ds^2=-U (r)\,dt^2 + U^{-1} (r)\,dr^2 + R^2 (r)\,d\Omega^2,\quad\phi=\phi (r)
\ee%
where $d\Omega^2$ is the metric element of the two-sphere $S^2$.

Finding exact solutions of the field equations stemming from the action (\ref{action})
is a very difficult task  even for simple forms  of  the potential $V$.   To solve the fields  equation 
we use the solution-generating technique  developed  in Ref.~\cite{Cadoni:2011nq} 
to find asymptotically AdS solutions once the scalar field profile $\phi=\phi(r)$ is given.
Let
\be%
R=e^{\int{}Y},\quad u=UR^2\label{nv},
\ee%
then the field equations take the simple form:
\bea%
Y'+Y^2&=&- (\phi')^2,\lb{riccati}\\
(u\phi')'&=&\frac{1}{4}\frac{\partial{V}}{\partial\phi}e^{2\int{}Y},\lb{sfe}\\
u''-4 (uY)'&=&-2,\lb{jq1}\\
u'' &=&2- 2Ve^{2\int{}Y}.\label{z3}
\eea%

Note that Eqs. (\ref{riccati}) and (\ref{jq1}) are universal, i.e.\ they do not depend on the potential.
We start with a given scalar field profile $\phi(r)$ and solve Eq. (\ref{riccati}) which is an
example of the Riccati equation in $Y$. Once $Y$
is known, we can  easily integrate the linear equation in $u$ (\ref{jq1}) to obtain
\be\lb{lsq}
u= R^4\left[-\int{}dr\left(\frac{2r+C_1}{R^4}\right)+C_2\right],
\ee%
where  $C_{1,2}$ are integration constants. 

The last step is to determinate the potential using Eq.~(\ref{z3})
\be\lb{ghi}
V=\frac{1}{R^{2}}\left(1-\frac{u''}{2}\right).
\ee%

This is a very efficient solving method, very useful in the holographic context,
allowing to find exact solutions of Einstein-scalar gravity in 
which the potential is not an input  but an output of the theory.

\section{The JNWW solutions\label{sec:3}}

The  parametrization (\ref{nv}) allows a simple rederivation of the solutions to the field
equations when $V=0$, i.e.\ the JNWW solutions.
Eq. (\ref{z3})  gives  $u$ as a quadratic function of $r$,   Eqs. (\ref{sfe}) and (\ref{jq1})  give  $\phi(r)$ and $R(r)$,
\be\lb{sol1}
U=\left(1-\frac{r_0}{r}\right)^{2w-1},\quad
R^2= r^2\left(1-\frac{r_0}{r}\right)^{2 (1-w)},\quad
\phi= -\gamma\ln\left(1-\frac{r_0}{r}\right)+\phi_0,
\ee%
whereas the  Riccati equation simply constrains the parameters, $w-w^2=\gamma^2$, therefore $0\leq w\leq1$.
Actually, the range of $w$ can be restricted to $1/2\leq w\leq1$ because of the invariance of the solution
under $w\to1-w$ together with a coordinate translation. Note that $w=1$ looks like the Schwarzschild solution with
a constant scalar field. Without loss of generality we can choose $\phi_0=0$.

According to old no-hair theorems,  this solution is not a BH
but interpolates between the Minkowski spacetime at $r=\infty$ and  a naked singularity at  $r=r_0$ (or $r=0$).
Near to the singularity the solution has a scaling behavior typical of  hyperscaling violation~\cite{Cadoni:2012uf}.

The  mass of this solution is $M=8\pi(2w-1)r_0$, hence we can have
a solution with zero or positive mass even in the presence of a naked singularity.
In particular for $w=1/2$ we have $M=0$, a degeneracy of the Minkowski vacuum.

 Note also that the JNWW solution appears as the zero-charge limit of charged dilatonic BHs.

\section{Asymptotic behavior of the scalar field and of the potential\label{sec:4}}

We look for  AF BH solutions sourced by  scalar field, which  decay as $1/r$
and we assume that $\phi=0$ corresponds to the Minkowki vacuum and at the same time it is an extremum of the potential with zero mass:
$V(0)=V'(0)=V''(0)=0$.

These conditions imply that near $\phi=0$ the potential behaves as $V(\phi)\sim\phi^n$ with $n\ge3$.
The corresponding asymptotic behavior for  the scalar is determined by imposing in the field
equations the boundary conditions  at $r\to\infty$, i.e.\ $u=r^2$ and $R=r$. For $n=5$ we get $\phi=\beta/r+\mathcal{O} (1/r^2)$,
and hence, an harmonic decay of the scalar field requires a  quintic behavior for the potential.

\section{Black hole solutions\label{sec:5}}

Let us now use the solution-generating method of Sect.~\ref{sec:2}.
Starting with the JNWW  scalar profile 
$\phi= -\gamma\ln\left(1-r_0/r\right)$,
the  Riccati equation gives the same metric function $R$ as in Eq. (\ref{sol1}).

Let $X\equiv1-r_0/r$. We get three different class of solutions 
\bea\lb{bhole1}
U (r) &=& X^{2w-1}\left[1-\Lambda\ (r^2+ (4w-3) rr_0 + (2w-1) (4w-3)
r_0^2)\right]
+\frac{\Lambda{}r^2}{X^{2(w-1)}}\\
U (r) &=&\frac{r^2}{r_{0}^{2}}X\left[\left(1+ r_{0}^{2}\Lambda\right)
X -2r_0^2\Lambda\ln{X}
+\left(1- r_{0}^{2}\Lambda\right) X^{-1}-2\right],\\
U (r)
&=&\frac{r^2}{r_0^2}X^{1/2}\left[\left(1+\frac{r_{0}^{2}\Lambda}{2}\right)
X^2 -2\left(1+r_{0}^{2}\Lambda\right) X+r_{0}^{2}\Lambda\ln{X}+1+
\frac{3r_{0}^{2}\Lambda}{2}\right]
\eea%
respectively for $1/2<w<1$, ($w\neq3/4$), $w=1/2$ and $w=3/4$.
In the previous equations, the $\Lambda$'s are appropriate (and different) 
combinations of the integration constants and the parameters chosen in order to have AF solutions.
The corresponding potentials are
\bea\lb{pot1}
V (\phi) &=&4\Lambda\left[w (4w-1)\sinh\frac{(2w-2)\phi}{\gamma}+8\gamma^2\sinh\frac{(2w-1)\phi}{\gamma}+\right.\nonumber\\
&&\qquad+ (1-w) (3 -4w)\left.\sinh\frac{2w\phi}{\gamma}\right],\\
V (\phi) &=&4
\Lambda\left[3\sinh2\phi-2\phi\left(\cosh2\phi+2\right)\right],\\
V (\phi) &=&
\Lambda\left(8\sqrt3\phi\cosh\frac{2\phi}{\sqrt3}-9\sinh\frac{2\phi}{\sqrt3}-\sinh2\sqrt3\phi\right)
\eea%

These solutions describe a one parameter family (the  scalar charge is not an independent parameter)
 of AF BHs with a curvature singularity at $r=r_0$ (or $r=0$) 
and a regular event horizon at $r=r_h$ sourced by a scalar field behaving asymptotically as $1/r$.
{\color{red}.} Near to the singularity the solution have the same 
scaling behavior of the JNWW solutions. As expected, near $\phi=0$ the potential has always a quintic behavior. 
The existence of these BH solutions represents a way to circumvent old and new no-hair theorems:
in fact the potential $V$ is not semipositive definite, it has an inflection point at $\phi=0$
and it is unlimited from below. Moreover, the ADM mass is not semipositive definite and hence the PET is violated.

\section{Black hole thermodynamics\label{sec:6}}

The scalar charge $\sigma$ is not independent from the mass but determined by the BH mass $M$, 
implying the absence of an associate thermodynamical potential. The  first principle has therefore the form  
$dM=T\,dS$, where  the temperature $T$ and the entropy $S$ are given by the usual  expressions $T=\frac{U'}{4\pi}|_{r=r_h}$,
$S=16\pi^2 R^2|_{r=r_h}$.

For $w=1/2$ we have 
\be\lb{ts}
T (\om)=\frac{\sqrt{\Lambda}}{4\pi\sqrt{\l}}\left[2\left(1-\frac{2}{\om}\right)\ln(1-\om)-4\right]\lb{pou},\quad
S (\om)=\frac{16\pi^2}{\Lambda\l}\left(\frac{1}{\om^2}-\frac{1}{\om}\right)
\ee%
where $\l$ is a function of $\om$ defined implicitly by
\[2(1-\om)\ln(1-\om)-\om^{2} (1+\l)+2\om=0.\]
We have an extremal low-mass state  with non vanishing mass, zero entropy and infinite temperature.
In the large mass (small temperature) limit we get the Schwarzschild behavior for the thermodynamical potentials:
\be%
M=\frac{2}{T},\quad S=\frac{1}{T^2},\quad F=M-TS=\frac{1}{T}.
\ee%

For $w=3/4$, $T$ and $S$ have a different behaviour, see details in Ref.~\cite{Cadoni:2015gfa}.
Both  the  low and large mass regimes have the Schwarzschild behavior. The extremal state has $M=S=0$ and $T=\infty$.
The thermodynamical behaviour of the  solutions with $1/2<w<3/4$ and $3/4<w<1$ are similar respectively to the cases $w=1/2$ and $w=3/4$.

\section{Concluding remarks\label{sec:7}}

AF BH solutions sourced by a scalar field with $1/r$ fall-off do exist but require a potential unlimited from below.
Because $\phi=0$ is an inflection point for $V$, the $\phi=0$ Schwarzschild BH is unstable.
We have observed that for large mass our BH is thermodynamical indistinguishable from the Schwarzschild one, while for $1/2\le w<3/4$ the low-mass regime is
drastically different.
Near to the $\phi=0$ Minkowski vacuum, the potential $V$ has a quintic behaviour, which means that the corresponding field theory is not renormalizable
and cannot be fundamental. However it could represent  an effective description arising from renormalization group flow.

\end{document}